\documentclass[fleqn,twoside]{article}
\usepackage{espcrc2}

\usepackage{graphicx}
\usepackage[figuresright]{rotating}

\newcommand{\be}{\begin{equation}}
\newcommand{\ee}{\end{equation}}
\newcommand{\bear}{\begin{eqnarray}}
\newcommand{\ear}{\end{eqnarray}}
\newcommand{\no}{\noindent}
\newcommand{\nn}{\nonumber}

\newcommand{{\tr}}{\rm tr}

\newcommand{\e}{{\rm e}}

\newcommand{\slD}{\raise.15ex\hbox{$/$}\kern-.57em\hbox{$D$}}

\newcommand{\tow}[2]{{\!\begin{array}{c}
{#1}\vspace*{-0.1cm}\\
{#2}\vspace*{-0.3cm}\end{array}\!}}

\newcommand{\AmS}{{\protect\the\textfont2
  A\kern-.1667em\lower.5ex\hbox{M}\kern-.125emS}}

\title{Determinant Calculations Using Random Walk Worldline Loops}

\author{Michael G. Schmidt\address[MCSD]{Institut f\"ur Theoretische Physik, Philosophenweg 16,
D-69120 Heidelberg, Germany}%
        \thanks{m.g.schmidt@thphys.uni-heidelberg.de}
               and
        Ion-Olimpiu Stamatescu\address[MCSD]{Forschungst\"atte der Evang. 
Studiengemeinschaft, Schmeilweg 5, D-69118, Heidelberg, 
Germany and Institut f\"ur Theoretische Physik, Philosophenweg 16,
D-69120 Heidelberg, Germany}%
\thanks{stamates@thphys.uni-heidelberg.de}}
       
\begin{document}

\begin{abstract}
We use statistical ensembles of  worldline
loops generated by random walk on hypercubic lattices to calculate matter
 determinants in background Yang-Mills fields.
\end{abstract}

\maketitle

 {\bf Introduction } 
The World Line Formalism (\cite{1} - see also \cite{3} and references therein)
permits a transparent discussion of matter
 determinants and the related 1-loop effective action in some background. We here consider a lattice implementation of this method.
Consider the bosonic determinant in
a gauge field background
\bear
\Gamma(A)\!\!&\!\!&\!= -\log\det ({\cal D}^2 + m^2)
          =   -\tr\log ({\cal D}^2 + m^2)   \nonumber \\
\!\!&\!\!&\!\!=\!\int^\infty_0\frac{dT}{T}\int d^Dx\int _{x(0)=x(T)=x}
[{ D}x]\, \nn \\
\times \!\!&\!\!&\!\!\tr_P\exp\left\{-\int^T_0d\tau\left(\frac{\dot x^2}{4}+
ig{\bf A}_\mu\dot x_\mu+m^2
\right)\right\} \nn
\ear
The (Euclidean) worldline sum  above is
over all closed paths ($\tr_P$ indicates path ordering).
 We use Random Walk (RW) paths on a D-dimensional hypercubic 
lattice $\Lambda_D$ \cite{12} and discretize $T$ using 
$\Delta \tau={T}/{L}={a^2}/{2D}$,
then:
\be
\Gamma^{discr}(U)=\sum_{x\in \Lambda_D} {\cal L}(U,x),
  \label{e.wlal}
\ee
\vspace{-0.5cm}
\bear
{\cal L}(U,x)=\sum_L\frac{1}{L}e^{-\frac{m^2 L}{2D}}
\left(\frac{1}{2D}\right)^L \! \sum_{\{\omega_L(x)\}}\!
\tr 
\! \prod_{l \in \omega_L(x)}\! U_l 
\nn
\ear
Dimensionfull quantities are understood as given in lattice units $a$. 
 $\{\omega_L(x)\}$ is the set of all closed lattice paths with length
$|\omega_L|=L$  obtained by RW starting and ending at $x$.
The free RW ``measure" $(2D)^{-L}$
is  implemented by the
actual procedure,  the mass dependence
$e^{-m^2T}$  is kept explicit. $U_l$ are link variables. For detail see \cite{rw}.

The expansion according to the loop lengths (\ref{e.wlal})
is equivalent to the
   usual hopping parameter expansion for the logarithm of the determinant  
(see, e.g., \cite{12}, \cite{kawa},
\cite{stadet}), e.g., 
in the bosonic case:
\bear
\kappa^L =(m^2+ 2 D)^{-L} \tow{\simeq}{m^2 \ll 2D} (2D)^{-L} \e^{- 
m^2 \frac{L}{2D}} \nn
\ear
The number of different loops of length $L$ increases exponentially with 
 $L$,
therefore one usually  stops the
expansion at a rather low order. Since 
loops  $L$ are typically relevant
 at  scale $\sqrt{L}$, we  then may completely miss  physical effects at 
 large scales  (or need to use rough lattices). 
 Our procedure amounts  instead to
sampling loops statistically in a wide interval of lengths  
(see also \cite{phil}).  By construction, the loops appear with the 
correct probabilities. We thus trade the systematic
errors 
for statistical ones, hoping to reproduce also effects
at ``large" scale,
albeit within statistical uncertainties. 
For a  quantum gauge field vacuum  the (quenched) vacuum energy
obtains by averaging (\ref{e.wlal}) with the  Yang-Mills  action. 

\begin{figure}[htb]
\vspace{3.2cm}
\includegraphics{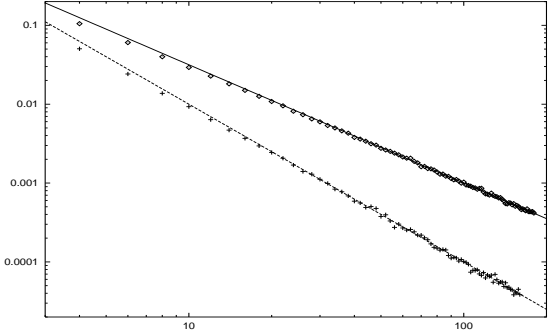}
\caption{
RW $\nu_L$ (\ref{e.freq}) vs $L$
in $D=3$ (diamonds)
and $4$  (crosses). 
Straight lines represent  $L^{-D/2}$.
}
\label{f.free}
\vspace{-0.3cm}
\end{figure}

 {\bf The lattice loop ensemble }  
The  factor $(2D)^{-L}$ in  (\ref{e.wlal})
is reproduced by RW and we only need to weight the loop
contributions $\tr\left( \prod_{l \in \omega_L} U_l \right)$ with 
$L^{-1} \e^{-m^2 \frac{L}{2D}}$. We specify a maximal number of trials ${\cal N}_L$  for each given number of steps $L$. We
first generate an $\{\omega(x_\mu = 1)\}$ ensemble of loops by starting 
 the random walk at the point
 $x_{\mu} = 1$ of
the periodic lattice and collecting the loops which after $L$ steps
have returned to $x_{\mu} = 1$, independently on  self-crossings or
retracing. Using this $\{\omega(x_\mu = 1)\}$ ensemble 
we define ${\cal L}(A,1)$. If we want to
 calculate ${\cal L}(A,x)$ 
we shift  the $\{\omega(x_\mu = 1)\}$
ensemble by $x$ as a whole. The full loop
ensemble --  the $x$-summation $\sum_{x\in \Lambda_D}$ in
(\ref{e.wlal}) -- is approximated by
considering  sufficiently many random translations of each of the
loops of the  $\{\omega(x_\mu = 1)\}$ ensemble
(this reproduces the correct 
 multiplicities in the $\kappa$ - expansion \cite{stadet}). 
We have for the  loop frequencies:
\be
\nu_L = \frac{2\,n_L}{(2\pi/D)^{D/2}
\, {\cal N}_L } \tow{\ =\ }{\ L \gg 1\ } L^{-D/2}.
\label{e.freq}
\ee 
with $n_L$ the number of loops generated at $L$. The factor 2 appears because 
$L$ is even. The second equality in (\ref{e.freq}) is the well known result for unbiased RW \cite{12}. Since the number of
loops decreases with $L$, if large loops are
 important we may have large statistical errors. One 
improves on this by using  larger ${\cal N}_L $ at high $L$, to produce 
more loops there (using $\nu_L$ automatically renormalizes their
contributions). We work in $D=3(4)$ and use  an $\{\omega(x_\mu = 1)\}$
ensemble of up to 600000(250000) loops on lattices of $N_x  N_y
  N_z   N_t$ with $N_{\mu}\, = \,12,\, 16,\, 24$ and 32, p.b.c.
See Fig. \ref{f.free}.
%

{\bf Special configurations } 
For a  test  we use:\par
\begin{figure}[htb]
\vspace{8.0cm}\includegraphics{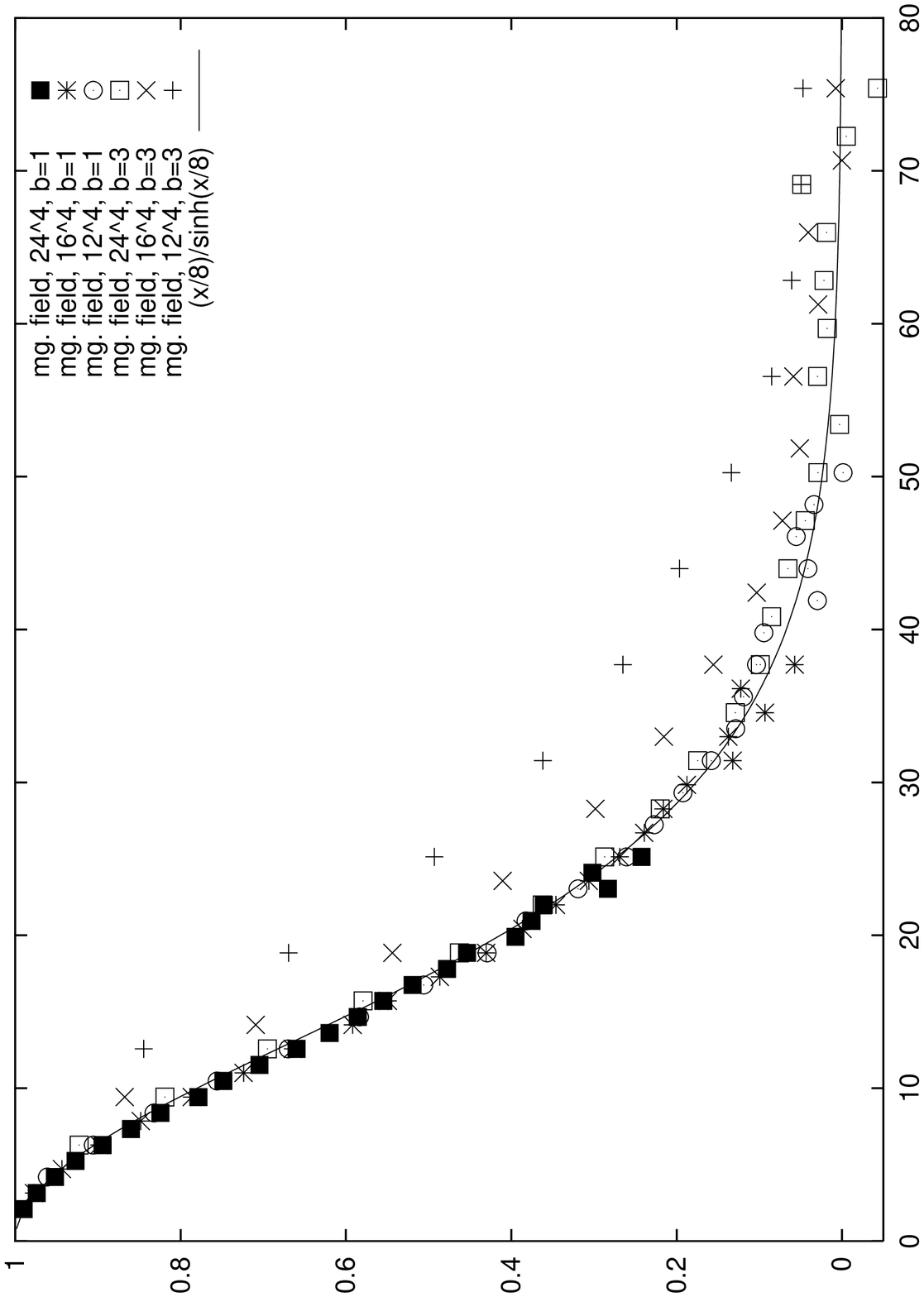}
\includegraphics{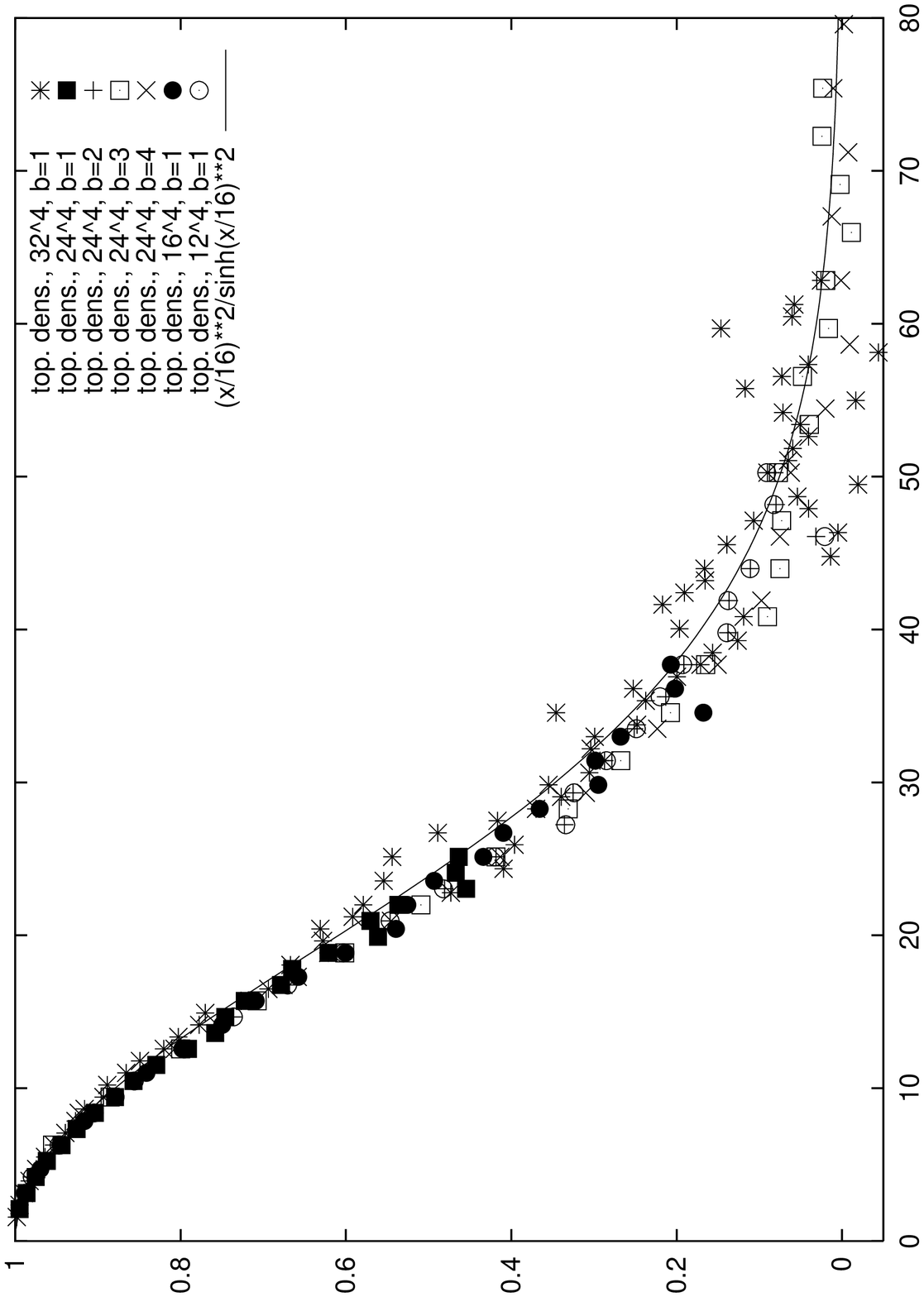}
\caption{
$F(L,b)$ vs $x \equiv bL$ 
in $D=4$,
for  U(1) (\ref{e.mgf1}) ($b$ in units of $4\pi/N$, upper  
plot) and SU(2) case (\ref{e.to1}) ($b$ in units of $2\pi / N$,
lower plot). The lines represent the continuum results (\ref{e.fc}), (\ref{e.tc}).}
\label{f.tcd}
\vspace{-0.5cm}
\end{figure}

\no {\it a) Homogeneous field}\hspace{0.1cm}
In continuum  \cite{4}:
\bear
{\cal L}(b_{+},b_{-}) &=& \frac{1}{(4 \pi)^{D/2}} \int_0^{\infty} \frac{dT}{ T} T^{-D/2}
\e^{-m^2 T} \nn \\
&\times&\frac{b_{+}T}{\sinh(b_{+}T)} \frac{b_{-}T}{\sinh(b_{-}T)},
\label{e.hfc1}
\ear
\be
b_{\pm}^2 = \frac{1}{2}\left[ {\vec E}^2 + {\vec B}^2 \pm
\sqrt{({\vec E}^2 + {\vec B}^2)^2 - 4 ({\vec E}{\vec B})^2} \right]
\label{e.hfc2}
\ee
(with 
subtraction at $T=0$). With  
\be
b_{\mu}= 2 \pi k_{\mu} /{{\hat N}_{\mu}}, \ \ k_{\mu}:\ {\rm integer},
\label{e.mgf2}
\ee
\be
U_{n, {\hat 1}} = \e^{ i n_y b_y}, \ 
U_{n, {\hat 2}} = \e^{- i n_x b_x}, \ 1\leq n_{\mu} \leq {\hat N}_{\mu}
\label{e.mgf1}
\ee
with ${\hat N}_{\mu} = N_{\mu}$ (all other links  1) realizes
a constant, Abelian  magnetic field $b=b_{+}=b_x+b_y,\ b_-=0$. 
Alternatively, with ${\hat N}_{\mu} = N_\mu = N$,
\be
U_{n, {\hat 1}} = \e^{ i \sigma_3  
n_y b_y},\  
U_{n, {\hat 3}} = \e^{- i \sigma_3 n_t b_t},\   
1\leq n_{\mu}\leq N
\label{e.to1}
\ee
(all other links  1)  realizes an SU(2) configuration of constant 
topological charge density $q= \frac{b^2}{8 \pi^2}$ (here $k_{y,t} = k;\ \ b_{\pm} = b =b_x=b_t$).
We calculate 
\be
F(L,b) = \frac{1}{n_L}
\sum_{i=1}^{n_L}\,\tr
\prod_{l \in \omega_L^i} U_l .
\ee
From (\ref{e.hfc1},\ref{e.hfc2})  this should be
\be
F_{\rm mag.}(L,b) = {u}/{\sinh u},\ \  u = {b L}/{2D},
\label{e.fc}
\ee
\be
F_{\rm top.}(L,b) = {u^2}/{\sinh^2u}, \ \  u= {bL}/{4D}
\label{e.tc}
\ee
for the first (\ref{e.mgf1}) and second (\ref{e.to1}) case, respectively.
For
 small fields the points scale well with the field and  lattice size  
 and 
reproduce the continuum results (\ref{e.fc}) or (\ref{e.tc}).
 For
large fields lattice artifacts are seen on small lattices. See Fig. \ref{f.tcd}.

\begin{figure}[htb]
\vspace{8.0cm}\includegraphics{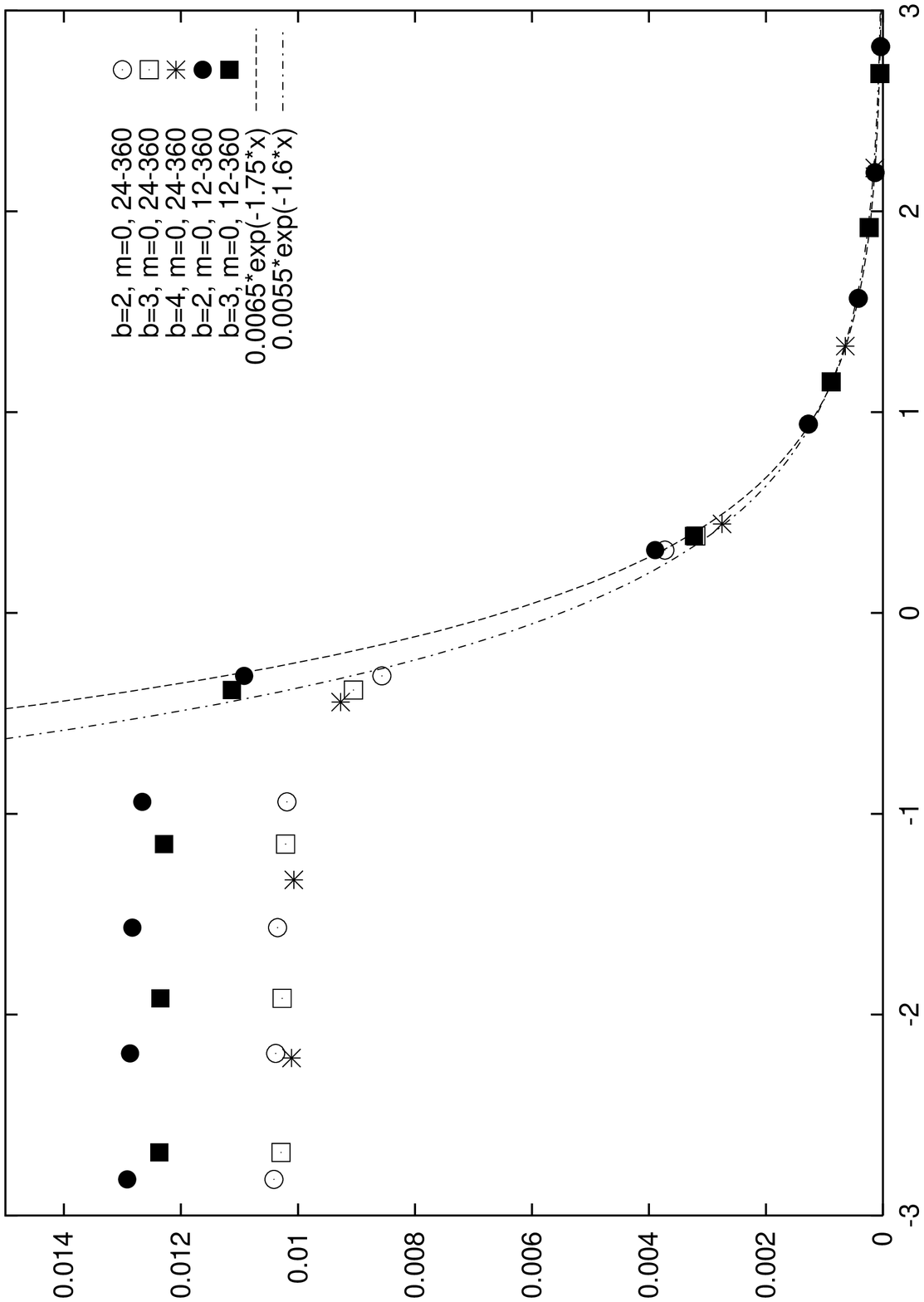}
\includegraphics{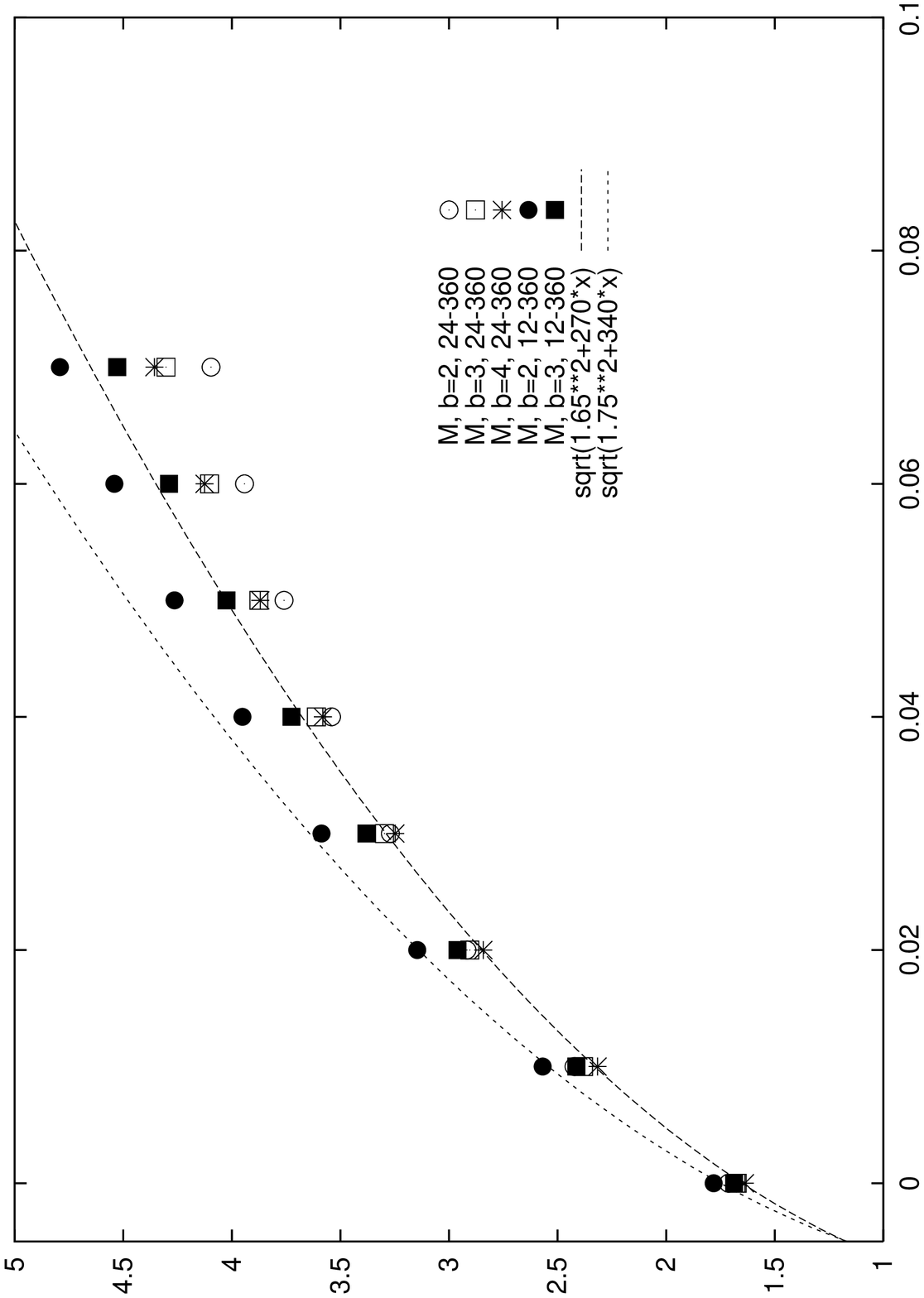}
\caption{
 $f(\lambda_1,\lambda_2,\mu,\xi)$
(\ref{e.scal1}) vs $x\equiv \xi$ at $\mu=0$ (upper plot), and 
 M vs $x \equiv \mu^2$ (lower plot), for various $b$ (in units of $\pi/16$) and $\lambda_1\,-\,\lambda_2$ intervals.}
\label{f.hmb1}
\vspace{-0.4cm}
\end{figure}

\no {\it b) Magnetic field in half space}\hspace{0.1cm}
Consider  a magnetic field  (\ref{e.mgf1}) in $D=3$
pointing into  $z$-direction 
 in half space $1 \leq x \leq {\hat N}_x = N_x/2$ 
\cite{11}. 
Then:
\bear
{\cal L}(L_1,L_2,b,m,x) = \sum_{L=L_1}^{L_2} \frac{1}{L} e^{-\frac{m^2 L}{2D}}\nu_L F(L,b,x) ,
\nn
\ear
\vspace{-0.4cm}
\be
F(L,b,x) = \frac{1}{n_L} 
\sum_{i=1}^{n_L}\,\tr
\prod_{l \in {\tilde \omega}_L^i(x)} U_l 
\ee
The lattice is $64\times 32^2, \ m\geq 0,\ {\tilde \omega}^i(x)$ are centered at $x$.
On physical grounds we expect
for ${\cal L}$ \cite{11}
\bear
{\cal L}\left(\frac{\lambda_1}{b},\frac{\lambda_2}{b},b,\mu \sqrt{b},
\xi\sqrt{b}+x_0 \right) =
b^{-\frac{3}{2}} f(\lambda_1,\lambda_2,\mu,\xi)
 \nn
\ear
\be
f(0,\infty,\mu,\xi) \propto \e^{-M(\mu) \xi}\ \ \  {\rm for}\ \ \xi > 0
\label{e.scal1}
\ee
Here $x_0= \frac{N_x}{2} +0.5$  ($\xi = 0$) is the separation between  regions. We have systematic errors due to $0 < \lambda_{1,2} < \infty$ and
discretization artifacts.
The slope $M$ appears reasonably stable against these uncertainties, however.
See Fig. \ref{f.hmb1}.
We also observe scaling (for small $b$ and $m$) and smooth dependence on
$m$. We obtain $M(0)=1.7 \pm 0.2$, in apparent disagreement with \cite{11}.
See \cite{rw}.

{\bf Discussion }  The above results indicate that using statistical ensembles of RW loops
 allows one to calculate the matter free energy
 in gauge
field backgrounds where large loops may be relevant. This method could be applied to non-trivial configurations such as instantons,
sphalerons, bounce solutions. It also could be used to calculate
the quenched free energy in a quantum vacuum \cite{5}.
Non-zero spin can also be handled by adding a spin term in the worldline
Lagrangian containing the outer field \cite{3} (for fermions it is 
advantageous to use second order formalism -- suggestion by W. Wetzel). See also \cite{17}.
Of course one can use direct numerical methods for  determinants. The above method may, however, 
allow more insight into the relevant phenomena and scales involved.

{\bf Acknowledgment:} We should like to thank
Martin Reuter and Werner Wetzel for discussions.


\begin{thebibliography}{}
\bibitem{1} R. P. Feynman, Phys. Rev. {\bf 80} (195)) 440.
\bibitem{3} 
M. Reuter, M. G. Schmidt, C. Schubert, Ann. of Physics {\bf 259} (1997)
319.
\bibitem{12}
C. Itzykson, J. M. Drouffe, Statistical Field Theory (Cambridge University
Press).
\bibitem{rw} M. G. Schmidt and I.-O. Stamatescu, hep-lat/0201002.
\bibitem{kawa}
N. Kawamoto, Nucl. Phys. {\bf B90}[FS3] (1981) 617.
\bibitem{stadet}
I.-O. Stamatescu, Phys. Rev. {\bf D25} (1982) 1130.
\bibitem{phil}
T. Bakeyev and Ph. De Forcrand, hep-lat/0008006.
\bibitem{4}
W. Heisenberg, H. Euler, Z. Phys. {\bf 38} (1936) 714;
J. Schwinger, Phys. Rev. {\bf 82} (1951) 664.
\bibitem{11}
H. Gies, K. Langfeld, Nucl. Phys. {\bf B613} (2001) 353.
\bibitem{5}
D. Fliegner, M. G. Schmidt, C. Schubert, Z. Phys. {\bf C64} (1994);
D. Fliegner, P. Haberl, M. G. Schmidt, C. Schubert, Ann. Phys. {\bf 264}  
(1998) 51.
\bibitem{17} M. Reuter, M. G. Schmidt, I.-O. Stamatescu, W. Wetzel,
work in progress.

\end{thebibliography}
\end{document}